\documentclass[aps,prl,twocolumn,superscriptaddress]{revtex4}
\usepackage{graphicx}
\usepackage{latexsym}
\usepackage{amssymb}
\usepackage{amsmath}
\usepackage{amsfonts}
\usepackage{bm}
\usepackage{multirow}
\usepackage{color}

\newcommand{\dsZ}{\mathbb{Z}}
\newcommand{\ii}{\mathrm{i}}
\newcommand{\dd}{\mathrm{d}}
\newcommand{\Tr}{\mathop{\mathrm{Tr}}}

\newcommand{\vect}[1]{{\bm{#1}}}
\newcommand{\eqnref}[1]{Eq.\,\eqref{#1}}

\newcommand{\tabref}[1]{Tab.\,\ref{#1}}

\newcommand{\beq}{\begin{equation}}
\newcommand{\eeq}{\end{equation}}
\newcommand{\beqn}{\begin{eqnarray}}
\newcommand{\eeqn}{\end{eqnarray}}

\begin{document}

\title{Bridging Fermionic and Bosonic Short Range Entangled States}


\author{Yi-Zhuang You}

\affiliation{Department of physics, University of California,
Santa Barbara, CA 93106, USA}

\author{Zhen Bi}

\affiliation{Department of physics, University of California,
Santa Barbara, CA 93106, USA}

\author{Alex Rasmussen}

\affiliation{Department of physics, University of California,
Santa Barbara, CA 93106, USA}

\author{Meng Cheng}

\affiliation{Station Q, Microsoft Research, Santa Barbara, CA
93106, USA}

\author{Cenke Xu}

\affiliation{Department of physics, University of California,
Santa Barbara, CA 93106, USA}

\begin{abstract}

In this paper we construct bosonic short range entangled (SRE)
states in all spatial dimensions by coupling a $\dsZ_2$ gauge
field to fermionic SRE states with the {\it same symmetries}, and
driving the $\dsZ_2$ gauge field to its confined phase. We
demonstrate that this approach allows us to construct many
examples of bosonic SRE states, and we demonstrate that the
previous descriptions of bosonic SRE states such as the
semiclassical nonlinear sigma model field theory and the
Chern-Simons field theory can all be derived using the fermionic
SRE states.

\end{abstract}


\maketitle

\emph{Introduction} ---

A short range entangled (SRE) state is the ground state of a
quantum many-body system that does not have bulk ground state
degeneracy or topological entanglement entropy. However, these
states can still have stable nontrivial edge states. Some of the
SRE states need certain symmetry to protect the edge states, and
these SRE states are also called symmetry protected topological
(SPT) states. The most well-known SPT states include the Haldane
phase of spin-1 chain~\cite{haldane1,haldane2}, quantum spin Hall
insulator~\cite{kane2005a,kane2005b}, topological
insulator~\cite{fukane, moorebalents2007, roy2007}, and
topological superconductor such as Helium$^3$-B
phase~\cite{roy2008, Qi_PRL2009}. All the free fermion SPT states
have been well understood and classified in
Ref.~\onlinecite{ludwigclass1,ludwigclass2,kitaevclass}, and
recent studies suggest that interaction may not lead to new SRE
states, but it can reduce the classification of fermionic SRE
states~\cite{fidkowski1,fidkowski2,qiz8,zhangz8,levinguz8,yaoz8,chenhe3B,senthilhe3}.
Unlike fermionic systems, bosonic SPT states do need strong
interaction. Most bosonic SRE states can be classified by symmetry
group cohomology~\cite{wenspt,wenspt2}, Chern-Simons
theory~\cite{luashvin} and semiclassical nonlinear sigma
model~\cite{xuclass}.

In this work we demonstrate that there is a close relation between
fermionic and bosonic SRE states, more precisely many bosonic SRE
states can be constructed from fermionic SRE states with the same
symmetry. All fermion systems have at least a $\dsZ_2$ symmetry
$c_i \rightarrow - c_i$, where $c_i$ is a local fermion
annihilation operator, thus we can couple all fermion Hamiltonians
to a dynamical $\dsZ_2$ gauge field, and microscopically this
$\dsZ_2$ gauge field commutes with the actual physical symmetry of
the fermion system.  Once the $\dsZ_2$ gauge field is in its
confined phase, the fermionic degree of freedom no longer exists
in the spectrum of the Hamiltonian, and the system becomes a
bosonic system. However, in many cases, confinement of a gauge
field necessarily breaks certain symmetry of the system, thus we
have to be very careful. In both $2d$ and $3d$, a $\dsZ_2$ gauge
field has a confined phase and a deconfined phase. The deconfined
phase is characterized by topological excitations of the $\dsZ_2$
gauge field. In $2d$, the $\dsZ_2$ gauge field has a ``vison"
excitation, which corresponds to a $\pi$-flux seen by the matter
fields. In $3d$, the topological excitation is a ``vison loop",
which is a closed ring of $\pi$-flux. In $2d$/$3d$, when the
visons/vison loops proliferate (condense), the system enters the
confined phase, \emph{i.e.} fermions carrying $\dsZ_2$ gauge
charge cannot propagate freely in the bulk due to the phase
fluctuations induced by the vison/vison loop condensation.

However, when the $\dsZ_2$ gauge field is coupled to a fermionic
SRE state, the vison and vison loop often carry nontrivial quantum
numbers, or degenerate low-energy spectrum. In these cases, when
visons and vison loops condense, the condensate would not be a
fully gapped nondegenerate state that does not break any symmetry.
Also, sometimes visons in $2d$ would have a nontrivial statistics,
thus it cannot trivially condense. Thus only in certain specific
cases can we confine the fermionic SRE states and obtain a fully
gapped and symmetric bosonic state. Thus analysis of spectrum and
quantum number carried by the vison and vison loop is the key of
our study.

Our approach can also be viewed as a slave fermion construction of
bosonic SRE states, which has been considered in
Ref.~\onlinecite{yewen1,yewen2,lulee,Tarun_PRB2013,xu2dspt}.
However, in all these previous studies the gauge group associated
with the slave fermion is bigger than $\dsZ_2$, which means that
when the gauge fluctuation is ignored, at the mean field level the
slave fermion has a much larger symmetry than the boson system,
and the analysis of gauge confined phase is much more complicated.
In our case the gauge group is $\dsZ_2$, and since any fermion
system has this $\dsZ_2$ symmetry, the fermion SRE states would
have the same symmetry as the bosonic states after gauge
confinement. Thus in our case the nature of the confined phase can
be analyzed reliably, and it only depends on the properties of
visons and vison loops.

\emph{Construction of 3d bosonic SPT phases} ---

Let us take the 3d topological superconductor (TSC) phase with
time-reversal symmetry as an example. One example of such TSC is
the $^3$He B phase. Here instead of focusing on the real $^3$He
system, we are discussing a more general family of TSC phases
defined on a lattice that are topologically equivalent to
$^3$He-B. One typical Hamiltonian of such TSC defined on the cubic
lattice reads \beqn H = \sum_{\vect{k}} \chi_{- \vect{k}} \Big[ \sum_{i
= 1}^3 \Gamma^i\sin k_i  - \Gamma^4 \big(3 - m - \sum_{i = 1}^3
\cos k_i \big) \Big] \chi_{\vect{k}}.\eeqn Here $m$ plays the same role
as the chemical potential in real $^3$He system: $m = 0$ is the
trivial-TSC transition critical point. The time-reversal symmetry acts as
$\chi_{\vect{k}} \rightarrow \ii \Gamma^5 \chi_{ - \vect{k}}$. Close
to the trivial-TSC phase transition, in the continuum limit this
TSC phase can be described by the following universal real space
Hamiltonian: \beqn H_0 = \int d^3 x \sum_{a = 1}^n
\chi^\intercal_a \left(\mathrm{i} \Gamma^1
\partial_x + \mathrm{i} \Gamma^2
\partial_y + \mathrm{i} \Gamma^3 \partial_z + m\Gamma^4 \right) \chi_a,
\cr\cr \Gamma^1 = \sigma^{30}, \ \Gamma^2 = \sigma^{10}, \
\Gamma^3 = \sigma^{22}, \ \Gamma^4 = \sigma^{21}, \ \Gamma^5 =
\sigma^{23}, \label{he3}\eeqn where $\sigma^{ij} = \sigma^i
\otimes \sigma^j$ denotes the tensor product of Pauli matrices,
and $a = 1 \cdots n$ is the flavor index. This is a widely used
approximate form for this class of TSC (For example,
Ref.~\onlinecite{qianomaly,ludwiganomaly}). For each flavor index
$a$, $\chi_a$ is a four component Majorana fermion. In this
Hamiltonian $m > 0$ and $m < 0$ correspond to the TSC phase and
the trivial phase respectively. The time-reversal symmetry acts as
$\dsZ_2^T: \chi \rightarrow \mathrm{i}\Gamma^5 \chi$. Our
conclusion is that, when we couple $n$-copies of this TSC to the
same $\dsZ_2$ gauge field, the $\dsZ_2$ gauge field can have a
fully gapped nondegenerate confined phase {\it when and only when}
$n$ is an integer multiple of $8$. And when $n = 8$, the confined
phase is the $3d$ bosonic SPT state with
time-reversal symmetry first characterized in
Ref.~\onlinecite{senthilashvin}.

First of all, when $n = 1$, the vison loop must be gapless, and
the gaplessness is protected by time-reversal
symmetry~\cite{Qi_PRL2009}. On a vison line along $x$ direction,
there will be a pair of counter-propagating Majorana modes, so the
effective $1d$ Hamiltonian along the vison line reads (see
Appendix B for derivation): \beqn\label{eq: vison Heff} H_{1d,x} =
\int \mathrm{d}x \ \chi^\intercal \mathrm{i} \sigma^3
\partial_x \chi. \eeqn In this reduced $1d$ theory, time-reversal
symmetry acts as $\dsZ_2^T: \chi \rightarrow \mathrm{i}\sigma^2
\chi$. The only mass term $\chi^\intercal \sigma^2 \chi$ in this
vison line would break time-reversal symmetry, thus as long as
time-reversal is preserved, the vison line is always gapless. This
implies that when $n = 1$ the vison line definitely cannot drive
the system into a fully gapped state by proliferation without
breaking time-reversal.

For $n > 1$, the effective theory along the vison line becomes
\beqn H_{1d,x} = \int \mathrm{d}x \sum_{a=1}^n \ \chi^\intercal_a
\mathrm{i} \sigma^3
\partial_x \chi_a. \label{visonx}\eeqn Then for even integer $n$,
it appears that there is a time-reversal symmetric mass term
$\chi^\intercal_{a} \sigma^1 A_{ab} \chi_b $, where $A$ is an
antisymmetric matrix in the flavor space. In the bulk theory
\eqnref{he3}, this mass term can correspond to several terms such
as $\chi^\intercal_{a} \sigma^{13} A_{ab} \chi_{b}$ (see Appendix
B). However, none of these terms can gap out vison lines along all
directions. For example, for vison loops along $y$ direction, the
modes moving along $+y$ is an eigenstate of $\Gamma^2$ with
$\Gamma^2 = +1$, and modes moving along $-y$ direction have
eigenvalue $\Gamma^2 = -1$. Because $\sigma^{13}$ commutes with
$\Gamma^2=\sigma^{10}$, $\chi^\intercal_{a} \sigma^{13} A_{ab}
\chi_{b}$ can never back-scatter modes in the $y$ vison line. In
fact no flavor mixing time-reversal invariant fermion {\it
bilinear} terms in the bulk would gap out the vison lines along
all directions, while a $\dsZ_2$ gauge confined phase requires
dynamically condensing vison lines in all directions. Therefore
the fermion bilinear flavor mixing terms in the bulk do not allow
us to condense the vison lines in order to generate a fully gapped
symmetric bosonic state.

Since no fermion bilinear term can gap out all the vison loops, we
need to consider interaction effects. In
Ref.\,\onlinecite{fidkowski1,fidkowski2}, the authors studied the
interaction effect on \eqnref{visonx}, and the conclusion is that
for $n = 8$ there is an SO(7) invariant interaction term
$H_{1d,\text{int}} = \int dx \;V_{abcd} \chi^\intercal_a \sigma^2
\chi_b \chi^\intercal_c \sigma^2 \chi_d $ that can gap out the
$1d$ theory \eqnref{visonx} without generating nonzero expectation
value of any fermion bilinear operator, where $V_{abcd}$ is some
coefficient tensor specified in
Ref.\,\onlinecite{fidkowski1,fidkowski2}. The same field theory
analysis applies here: the effective interaction
$H_{1d,\text{int}}$ can gap out the $1d$ theory \eqnref{visonx}
along the vison loop without degeneracy. $H_{1d,\text{int}}$
corresponds\footnote{The fact that $\sigma^2$ on the vison line is
extended to $\Gamma^5$ in the bulk is explained in Appendix B.} to
the following term in the bulk: \beqn H_\text{int} = \int
\mathrm{d}^3x \;V_{abcd} \chi^\intercal_a \Gamma^5 \chi_b
\chi^\intercal_c \Gamma^5 \chi_d. \eeqn Since this term is
rotationally invariant, it will gap out vison lines along all
directions. Thus with $n = 8$, and with the interaction term
$H_\text{int}$ in the bulk, all vison loops can be gapped out
without breaking time-reversal symmetry, thus we can safely
condense the vison loops and drive the system into a fully gapped,
time-reversal invariant bosonic state. But this is only possible
when $n$ is an integral multiple of $8$. In the following
paragraphs we will argue that when $n = 8$ the confined bosonic
state is a bosonic SPT state.

Ref.~\cite{senthilashvin,xuclass} pointed out that this $3d$
bosonic SPT state can be described by a O(5) NLSM field theory
with a topological $\Theta-$term. Let us couple the 8 copies of
$^3$He B to a five-component unit vector $\vect{n}$: \beqn H = H_0
+ \int d^3x \ \sum_{j = 1}^5 n^j \chi^\intercal_a \Gamma^5
\gamma^j_{ab}\chi_b, \eeqn where $\gamma^j$ are five $8 \times 8$
symmetric matrices in the flavor space that satisfy $\{\gamma^i,
\gamma^j\} = 2\delta_{ij}$ (e.g. a particular choice could be
$\gamma^{i}=\sigma^{100},\sigma^{310},\sigma^{331},\sigma^{333},\sigma^{212}$).
Under time-reversal transformation, $\vect{n} \rightarrow -
\vect{n}$. Following the calculation in
Ref.~\onlinecite{abanov2000}, we can show that for the $^3$He B
phase with $m > 0$, after integrating out the fermions, the
effective field theory for the vector $\vect{n}$ contains a
topological $\Theta$-term at $\Theta = 2\pi$: \beqn S = \int
\mathrm{d}^3x \mathrm{d}\tau \ \frac{1}{g} (\partial_\mu
\vect{n})^2 + \frac{\mathrm{i} \Theta}{ \Omega_4 }
\epsilon_{abcde} n^a
\partial_x n^b \partial_y n^c
\partial_z n^d
\partial_\tau n^e, \label{o5nlsm} \eeqn where $\Omega_4$ is
the volume of a four dimensional sphere with unit radius.
\eqnref{o5nlsm} is precisely the field theory introduced in
Ref.~\onlinecite{senthilashvin,xuclass} to describe the $3d$
bosonic topological SC with time-reversal symmetry.

Using the field theory \eqnref{o5nlsm}, we can demonstrate that
the $2d$ boundary of this $3d$ bosonic SPT state could be a $2d$
$\dsZ_2$ topological order, whose {mutually semionic} excitations
$e$ and $m$ are both Kramers' doublet~\cite{senthilashvin} (The so
called $eTmT$ state)\footnote{The $\dsZ_2$ topological order at
the $2d$ boundary has nothing to do with the bulk $\dsZ_2$ gauge
field that we will confine by proliferating the vison loops.}.
Ref.~\onlinecite{chenhe3B,senthilhe3,metlitski_unpub} argued that
the boundary of 8 copies of $^3$He B is the (fermionized)~$eTmT$
state. For the sake of completeness, we will repeat this argument.
Based on the field theory \eqnref{o5nlsm}, the $e$ and $m$
excitations at the $2d$ boundary of the $3d$ bosonic SPT phase
correspond to the vortex of boson field $b_1 \sim n_1 +
\mathrm{i}n_2$, and vortex of $b_2 \sim n_3 + \mathrm{i} n_4$
respectively \footnote{assuming tentatively an enlarged
$U(1)\times U(1)$ symmetry for rotation of $(n_1, n_2)$ and $(n_3,
n_4)$ respectively}, which can be considered as surface
terminations of bulk vortex lines. By solving the Bogoliubov-de
Gennes equation with a vortex at the boundary, we can demonstrate
that there are four Majorana fermion zero modes located at each
vortex core. These four Majorana fermion zero modes can in total
generate four different states. Under interaction, time-reversal
symmetry~\footnote{Here the time-reversal symmetry is a modified
time-reversal symmetry defined in Ref.~\cite{senthilhe3}, which is
a product of ordinary time-reversal and a $\pi-$rotation of boson
field $b_1$ or $b_2$.} guarantees that these four states split
into two degenerate doublets with opposite fermion number parity.
Thus in the bulk each vortex line is effectively four copies of 1d
Kitaev's Majorana chain. Since we are in a $\dsZ_2$ gauge confined
phase, we are only allowed to consider states with even number of
fermions, thus after gauge projection, only one of the two
doublets survives, which according to the supplementary material
and Ref.~\onlinecite{senthilhe3} is a Kramers doublet. Also the
vortex of $b_1$ carries charge $\pm 1/2$ of $b_2$, and vortex of
$b_2$ carries $\pm 1/2$ charge of $b_1$, thus these two vortices
are both Kramers doublet, and they have mutual semion statistics.
This means that boundary of the confined phase is really the
$eTmT$ state.

Combining all the results together, we conclude that the $\dsZ_2$
confined phase of 8 copies of $^3$He B is really the bosonic SPT
phase with time-reversal symmetry. Furthermore, since this bosonic
SPT state has $\dsZ_2$ classification, it implies that two copies
of the bosonic state is trivial (which can be shown in our NLSM
field theory by directly coupling two copies of Eq.~\ref{o5nlsm}
together~\cite{xuclass}), which then implies that 16 multiples of
the $^3$He-B TSC is trivial under interaction. This conclusions is
consistent with the well-known $\dsZ_{16}$ classification of DIII
class fermionic SPT states\cite{chenhe3B,senthilhe3,youinversion}.

We can also give the 8 copies
of $^3$He B phase various flavor symmetries, and we can construct
many $3d$ bosonic SPT phases with symmetry that contains $\dsZ_2^T$
as a normal subgroup by confining the bulk $\dsZ_2$ gauge field.
Since all the free fermion SPT states in $3d$ require the
time-reversal symmetry, thus so far our approach does not allow us
to construct $3d$ bosonic SPT phases without $\dsZ_2^T$.

\emph{Construction of 2d bosonic SPT phases} ---

Now let us look at $2d$ examples. In $2d$ the simplest fermionic
SRE state is the $p+\mathrm{i}p$ topological superconductor (TSC)
that does not require any symmetry, and the simplest bosonic SRE
state is the so called ``$E_8$" state with chiral central charge
$c_- = 8$ at its boundary~\cite{e8, kitaev_talk2}. In the
following we will prove that if we couple $n$ copies of
$p+\mathrm{i}p$ TSC to a $\dsZ_2$ gauge field, the $\dsZ_2$ gauge
field can confine to a gapped bosonic state when and only when $n$
is a integral multiple of 16. And when $n = 16$, the confined
phase is precisely the bosonic $E_8$ SRE
state~\cite{Plamadeala_PRB2013}. First of all, when $n = 1$, the
vison of the $\dsZ_2$ gauge field carries a Majorana fermion zero
mode, which grants the vison a nonabelian statistics, thus when
$n=1$ (and generally for odd integer $n$) the $\dsZ_2$ gauge field
cannot enter its confined phase by condensing the vison. When $n$
is even, $n$-copies of $p+\mathrm{i}p$ TSC is equivalent to an
integer quantum Hall (IQH) state with Hall conductivity $\nu =
n/2$, thus a vison (half flux quantum) would carry charge $n/4$,
and has statistics angle $\pi n/8$ under exchange. Thus the
smallest $n$ that makes vison a boson is $16$, and when $n = 16$,
the $\dsZ_2$ gauge field can enter a confined phase by condensing
the bosonic vison.

The vison condensation can be formulated by the Chern-Simons
theory.\cite{Cheng:2014cr} Let us start from the Chern-Simons
description for $n$-copies of $p+\mathrm{i}p$ TSC with even
$n=2\nu$ (\emph{i.e.} $\nu$ layers of IQH), and couple the fermion
currents $da^I$ ($I=1,\cdots,\nu$) to the $\dsZ_2$ gauge field.
The Lagrangian density can be written as
\begin{equation}\label{eq: IQH gauge}
\mathcal{L}=\sum_{I}\frac{1}{4\pi}a^I\wedge
da^I+\sum_{I}\frac{1}{2\pi}A\wedge da^I+\frac{1}{\pi}A\wedge
d\tilde{A}.
\end{equation}
Here the $\dsZ_2$ gauge theory is described by the mutual
Chern-Simon theory of two gauge fields $A$ and $\tilde{A}$ with
the $K$-matrix $[\begin{smallmatrix}0 & 2\\
2&0\end{smallmatrix}]$, which is a standard representation of
$\dsZ_2$ gauge theory, for instance see
Ref.~\cite{kouwen,xutriangle}. $\tilde{A}$ can be considered as a
Higgs field that Higgs the U(1) gauge structure of $A$ down to
$\dsZ_2$. The field $A$ couples to the fermion current
$j^I=\star\dd a^I$ with equal charge, and the field $\tilde{A}$
couples to the vison current in the $\dsZ_2$ gauge theory. The
field $A$ can be treated as a Lagrangian multiplier and integrated
out first, which leads to the constraint $\sum_I
a^I+2\tilde{A}=0$. This constraint can be solved by the following
reparameterization
\begin{equation}\label{eq: V for IQH}
\begin{split}
& a^1=\tilde{a}^1, a^{\nu-1} = \tilde{a}^{\nu}+\tilde{a}^{\nu-1}-\tilde{a}^{\nu-2}, a^{\nu}=\tilde{a}^{\nu}-\tilde{a}^{\nu-1},\\
& a^{I}=\tilde{a}^{I}-\tilde{a}^{I-1} (\text{for
}I=2,\cdots,\nu-2), \tilde{A}=-\tilde{a}^{\nu}.
\end{split}
\end{equation}
Substituting \eqnref{eq: V for IQH} into \eqnref{eq: IQH gauge},
we arrive at a bosonic theory in terms of the new set of gauge
fields $\tilde{a}^I$, as
$\mathcal{L}=\sum_{I,J}\frac{1}{4\pi}K_{IJ}^{SO(n)}\tilde{a}^I\wedge
d\tilde{a}^J$, where $K^{SO(n)}$ is the Cartan matrix of the
$\mathfrak{so}(n)$ Lie algebra(for even $n>2$).  For $n=16$, the
$K$-matrix reads
\begin{equation}
K^{SO(16)}=
\left[\begin{smallmatrix}
2 & -1 & 0 & 0 & 0 & 0 & 0 & 0 \\
 -1 & 2 & -1 & 0 & 0 & 0 & 0 & 0 \\
 0 & -1 & 2 & -1 & 0 & 0 & 0 & 0 \\
 0 & 0 & -1 & 2 & -1 & 0 & 0 & 0 \\
 0 & 0 & 0 & -1 & 2 & -1 & 0 & 0 \\
 0 & 0 & 0 & 0 & -1 & 2 & -1 & -1 \\
 0 & 0 & 0 & 0 & 0 & -1 & 2 & 0 \\
 0 & 0 & 0 & 0 & 0 & -1 & 0 & 2
\end{smallmatrix}\right],
\end{equation}
which gives the $SO(16)_1$ Chern-Simons theory.
We now extend $K^{SO(16)}$ by a block of trivial boson, given by
the $K$-matrix $\sigma^1$~\cite{Cano2014},  and define
$K^{\text{ext}}=K^{SO(16)}\oplus \sigma^1$. One finds {a}
transform $W$, with $\det W=1$, given by
\begin{equation}
W^{-1}=
\left[\begin{smallmatrix}
 1 & 0 & 0 & 0 & 0 & 0 & 1 & -1 & 0 & 0 \\
 0 & 1 & 0 & 0 & 0 & 0 & 2 & -2 & 0 & 0 \\
 0 & 0 & 1 & 0 & 0 & 0 & 3 & -3 & 0 & 0 \\
 0 & 0 & 0 & 1 & 0 & 0 & 4 & -4 & 0 & 0 \\
 0 & 0 & 0 & 0 & 1 & 0 & 5 & -5 & 0 & 0 \\
 0 & 0 & 0 & 0 & 1 & 0 & 3 & -4 & -1 & 0 \\
 0 & 0 & 0 & 0 & 1 & 0 & 1 & -2 & 0 & 0 \\
 0 & 0 & 0 & 0 & 1 & -1 & 3 & -2 & 0 & 0 \\
 0 & 0 & 0 & 0 & 1 & -1 & 0 & 0 & -1 & 0 \\
 0 & 0 & 0 & 0 & 0 & 0 & 0 & 1 & 1 & -1
\end{smallmatrix}\right],
\end{equation}
such that
\begin{equation}
W^{\intercal}K^{\text{ext}}W=
\left[\begin{smallmatrix}
 2 & -1 & 0 & 0 & 0 & 0 & 0 & 0 & 0 & 0 \\
 -1 & 2 & -1 & 0 & 0 & 0 & 0 & 0 & 0 & 0 \\
 0 & -1 & 2 & -1 & 0 & 0 & 0 & 0 & 0 & 0 \\
 0 & 0 & -1 & 2 & -1 & 0 & 0 & 0 & 0 & 0 \\
 0 & 0 & 0 & -1 & 2 & -1 & 0 & -1 & 0 & 2 \\
 0 & 0 & 0 & 0 & -1 & 2 & -1 & 0 & 0 & -1 \\
 0 & 0 & 0 & 0 & 0 & -1 & 2 & 0 & 0 & -1 \\
 0 & 0 & 0 & 0 & -1 & 0 & 0 & 2 & 0 & -2 \\
 0 & 0 & 0 & 0 & 0 & 0 & 0 & 0 & 0 & 2 \\
 0 & 0 & 0 & 0 & 2 & -1 & -1 & -2 & 2 & 0
\end{smallmatrix}\right],
\end{equation}
The last $2\times 2$ block describes a $\dsZ_2$ topological order.
The fermion excitations of this $K$-matrix corresponds to the
original fermion in the $p+\mathrm{i}p$ TSC. The vison couples to
the last gauge field, \emph{i.e.} it corresponds to the charge
vector $(0,0,0,0,0,0,0,0,0,1)$, and is a boson ready to condense.
Thus after the vison condensation, the $\dsZ_2$ topological order
is destroyed and the original fermion is confined. The $K$-matrix
is left with the upper $8\times8$ block, which is exactly the
Cartan matrix of the ${E}_8$ Lie algebra. Since all the charge
vectors of the upper $8 \times 8$ block are self-bosons, and they
are bosons relative to the vison, these charge vectors are
unaffected by the vison condensate. Thus we have shown by explicit
calculation that confining the fermions in 16-copies of
$p+\mathrm{i}p$ TSC leads to the $E_8$ bosonic SRE state.

Now let us investigate the $p\pm\mathrm{i}p$ TSC with a $\dsZ_2$
symmetry discussed in Ref.~\onlinecite{levinguz8}. In this system
the fermions with zero $\dsZ_2$ charge form {a} $p+\mathrm{i}p$
TSC, while fermions carrying $\dsZ_2$ charge form {a}
$p-\mathrm{i}p$ TSC. This $\dsZ_2$ global symmetry is different
from the $\dsZ_2$ gauge symmetry, since all the fermions in our
system carry $\dsZ_2$ gauge charge. For one copy of the
$p\pm\mathrm{i}p$ TSC coupled to the $\dsZ_2$ gauge field, the
vison carries two independent Majorana fermion zero modes $\chi_1$
and $\chi_2$, and the global $\dsZ_2$ symmetry acts $\dsZ_2 : \chi
\rightarrow \sigma^z \chi $. There is no nontrivial Hamiltonian
for these two Majorana fermion modes that preserves the $\dsZ_2$
symmetry, thus the spectrum of the vison is always two fold
degenerate, and hence condensing the vison will not lead to a
nondegenerate state.

Two copies of the $p\pm\mathrm{i}p$ TSC is formally equivalent to
a quantum spin Hall (QSH) insulator: fermions that carry global
$\dsZ_2$ charge $0$ and $1$ form $\nu = 1$ and $-1$ integer
quantum Hall states respectively. Then after coupling to the
$\dsZ_2$ gauge field, the vison would carry two complex localized
fermion modes $c_1$ and $c_2$, and a vison would carry charge $\pm
1/2$ of the $\dsZ_2$ global symmetry, which corresponds to $n_2 =
c^\dagger_2 c_2 = 1$, $0$ respectively (The $\dsZ_2$ symmetry in
our system is just the $\dsZ_2$ subgroup of the U(1) symmetry of
the $\nu = -1$ integer quantum Hall state, and it is known that a
vison, or a $\pi-$flux in a $\nu = -1$ integer quantum Hall state
carries $\pm 1/2$ charge, as was shown in
Ref.~\onlinecite{ranlee}). Thus the condensate of the vison always
spontaneously breaks the $\dsZ_2$ symmetry. This situation is very
similar to the case discussed in Ref.~\onlinecite{ranlee}. The
universality class of the confinement transition is the so-called
$3d$ XY$^\ast$ transition, namely at the quantum critical point
the $\dsZ_2$ symmetry order parameter has {an} anomalous dimension
$\eta \sim 1.49$~\cite{melko1,melko2}.

Eventually for four copies of this $p\pm\mathrm{i}p$ TSC, a vison carries
four complex fermion modes $c_{1A}$, $c_{1B}$, $c_{2A}$, $c_{2B}$.
The vison now can be a boson that does not carry any $\dsZ_2$ global
charge, for example the state with $n_{2A} = 1$ and $n_{2B} = 0$
is a $\dsZ_2$ charge neutral boson. Thus condensing this vison would
lead to a fully gapped nondegenerate bosonic state that preserves
the global $\dsZ_2$ symmetry.

Now let us couple four copies of the $p\pm\mathrm{i}p$ TSC to a
four-component unit vector $\vect{n}$:
\begin{equation}
\begin{split}
H =\int\mathrm{d}^2x\  &\chi^\intercal ( \mathrm{i}
\sigma^{3000}
\partial_x + \mathrm{i} \sigma^{1000} \partial_y + m \sigma^{2300}) \chi
\\
&+ \sum_{j = 1}^4 n^j \chi^\intercal \gamma^j \chi,
\end{split}
\end{equation}
with $\gamma^1 =
\sigma^{2100}$, $\gamma^2 = \sigma^{2221}$, $\gamma^3 =
\sigma^{2223}$, $\gamma^4 = \sigma^{2202}$.
The global $\dsZ_2$
symmetry acts as $\dsZ_2: \chi \rightarrow \sigma^{0300} \chi$, and
$\vect{n} \rightarrow - \vect{n}$. After integrating out the
fermions, the resulting theory is a $(2+1)d$ O(4) NLSM with a
topological $\Theta$-term at $\Theta = 2\pi$:
\begin{equation}
S =\int \mathrm{d}^2x\mathrm{d}\tau \frac{1}{g} (\partial_\mu
\vect{n})^2 + \frac{\mathrm{i}\Theta}{ \Omega_3}\epsilon_{abcd}
n^a \partial_x n^b
\partial_y n^c \partial_\tau n^d,
\end{equation}
where $\Omega_3=2\pi^2$ is the volume of a three dimensional
sphere with unit radius, and this is precisely the field theory
describing the $2d$ bosonic SPT phase with $\dsZ_2$ symmetry,
which was first studied in Ref.~\onlinecite{levingu}. This field
theory was studied in Ref.~\onlinecite{xusenthil,xuclass}.

Finally we condense the vison in this system to confine the
fermions. Similar to our previous $K$-matrix calculation, we
couple the four copies of $p\pm\mathrm{i}p$ TSC to the $\dsZ_2$
gauge field, as described by the Lagrangian density
\begin{equation}\label{eq: QSH gauge}
\mathcal{L}=\sum_{I,J}\frac{K_{IJ}^{\text{QSH}}}{4\pi}a^I\wedge
\dd a^J+\sum_{I}\frac{1}{2\pi}A\wedge \dd a^I+\frac{1}{\pi}A\wedge
\dd\tilde{A},
\end{equation}
where the matrix $K^{\text{QSH}}$ is diagonal with the diagonal
elements $(1,1,-1,-1)$. In the theory, the global $\dsZ_2$ symmetry
charge is given by the charge vector $q_{\dsZ_2}=(0,0,1,1)$.
Integrating out $A$ leads to the constraint
$\sum_{I}a^{I}+2\tilde{A}=0$, which can be solved by
\begin{equation}\label{eq: V for QSH}
\left[\begin{matrix}
a^1 \\ a^2 \\ a^3 \\ a^4 \\ \tilde{A}
\end{matrix}\right]=
\left[\begin{matrix}
 1 & 1 & -1 & 1 \\
 0 & 0 & 1 & 1 \\
 1 & 0 & -1 & 1 \\
 0 & -1 & 1 & -1 \\
 -1 & 0 & 0 & -1
\end{matrix}\right]
\left[\begin{matrix}
\tilde{a}^1 \\ \tilde{a}^2 \\ \tilde{a}^3 \\ \tilde{a}^4
\end{matrix}\right].
\end{equation}
Substituting \eqnref{eq: V for QSH} into \eqnref{eq: QSH gauge}
yields a Chern-Simons theory
$\mathcal{L}=\sum_{I,J}\frac{1}{4\pi}K_{IJ}^{\text{SPT}^*}\tilde{a}^I\wedge
d\tilde{a}^J$ with
\begin{equation}
K^{\text{SPT}^*}=
\left[\begin{matrix}
 0 & 1 & 0 & 0 \\
 1 & 0 & 0 & 0 \\
 0 & 0 & 0 & 2 \\
 0 & 0 & 2 & 0
\end{matrix}\right].
\end{equation}
Correspondingly, the global $\dsZ_2$ charge is transformed to
$\tilde{q}_{\dsZ_2}=W^{\intercal} q_{\dsZ_2}=(1,-1,0,0)$, with the
transformation matrix $W$ taken from the first 4 rows of the
matrix in \eqnref{eq: V for QSH}. In $K^{\text{SPT}^*}$, the lower
$2\times2$ block describes the $\dsZ_2$ topological order, which
contains the bosonic vison with neutral global $\dsZ_2$ charge (as
seen from $\tilde{q}_{\dsZ_2}$). As the vison condenses, the
$\dsZ_2$ topological order is removed, leaving the upper
$2\times2$ block, \emph{i.e.} the $\sigma^1$ matrix, as the
$K$-matrix describing a SRE bosonic state, with the global
$\dsZ_2$ charge $q=(1,-1)$ (as taken from $\tilde{q}_{\dsZ_2}$).
Such a $K$-matrix equipped with the $\dsZ_2$ symmetry
matches~\cite{luashvin} the Chern-Simons description of the
$\dsZ_2$ SPT state. Therefore after confining the fermions in four
copies of $p\pm\mathrm{i}p$ TSC, we obtain the bosonic SPT state
with $\dsZ_2$ global symmetry. This bosonic SPT state has $\dsZ_2$
classification~\cite{wenspt,levingu,xuclass}, which implies that 8
copies of the $p\pm\mathrm{i}p$ TSC with $\mathbb{Z}_2$ symmetry
is a trivial state, which is consistent with the well-known
$\dsZ_8$ classification of such $p\pm\mathrm{i}p$ TSC under
interaction~\cite{qiz8,yaoz8,zhangz8,levinguz8,youinversion}

Extra symmetries can be added to the four copies of $p \pm
\mathrm{i}p$ TSC discussed above, and other $2d$ bosonic TSC can
be constructed in the same way. Construction of $1d$ bosonic SPT
phases is much more obvious, which will be discussed in the
supplementary material.

\emph{Summary} ---

In this paper we demonstrate that many bosonic SRE phases can be
constructed by fermionic SRE phases with the same symmetry. The
fermionic SRE states and the $\dsZ_2$ gauge field can all be defined
on a lattice, thus our method has provided a projective
construction of the lattice wave function of these bosonic SRE
states. Also, our method provides a full lattice regularization of
the CS field theory~\cite{luashvin} and semiclassical NLSM field
theory~\cite{xuclass} description of bosonic SPT phases. However,
some bosonic SPT phases cannot be constructed using the method
discussed in the current paper, for example, there is one bosonic
SPT phase with $U(1)\rtimes \dsZ_2$ symmetry in $3d$, while there is
no free fermion SPT phase with the same symmetry. We will leave
the construction of these bosonic SPT phases to future study.

The authors are supported by the the David and Lucile Packard
Foundation and NSF Grant No. DMR-1151208.


\appendix
\section{Appendix A. Construction of $1d$ Bosonic SPT}
In this appendix, we construct the $1d$ Haldane phase using four
copies of Kitaev's chains with the time-reversal
symmetry$\dsZ_2^T$.
Let us start from the fermionic SPT phase composed of four copies
of Kitaev's chains coupled to a fluctuating three-component unit
vector $\vect{n}$:
\begin{equation}\label{eq: H 1d FSPT}
H=\chi^\intercal(\mathrm{i}\sigma^{100}\partial_x+m\sigma^{200})\chi+\sum_{j=1}^{3}n^j
\chi^\intercal\gamma^j\chi,
\end{equation}
with $\gamma^{1}=\sigma^{332}$, $\gamma^{2}=\sigma^{320}$,
$\gamma^{3}=\sigma^{312}$. The time reversal symmetry acts as
$\dsZ_2^T:\chi\to\sigma^{300}\chi$ and $\vect{n}\to-\vect{n}$
followed by the complex conjugation (denoted $\mathcal{K}$). Note
that the time reversal operator
$\mathcal{T}=\mathcal{K}\sigma^{300}$ behaves as $\mathcal{T}^2=1$
on the Majorana fermions $\chi$. After integrating out the
fermions, the resulting theory is a $(1+1)d$ O(3) NLSM with a
topological $\Theta$-term at $\Theta=2\pi$:
\begin{equation}
S =\int \mathrm{d}x\mathrm{d}\tau \frac{1}{g} (\partial_\mu
\vect{n})^2 + \frac{\mathrm{i}\Theta}{ \Omega_2}\epsilon_{abc} n^a
\partial_x n^b
\partial_\tau n^c,
\end{equation}
where $\Omega_2=4\pi$ is the volume of a two dimensional sphere
with unit radius, and this is precisely the field theory
describing the $1d$ bosonic SPT phase with $\dsZ_2^T$ symmetry,
\emph{i.e.} the Haldane phase of $1d$ spin
chain~\cite{haldane1,haldane2}.

Then we can couple the fermions to a $\dsZ_2$ gauge field, namely
we impose the following gauge constraint on every site:
$\chi_{i0}\chi_{i1}\chi_{i2}\chi_{i3} = 1$. The same gauge
constraint is imposed on the edge Majorana fermion zero modes. The
edge Majorana fermion zero modes may be arranged in a matrix as\cite{hermele}
\begin{equation}
F=\frac{1}{2}(\chi_{0}\sigma^0+ \mathrm{i}\chi_{1}\sigma^1+
\mathrm{i}\chi_{2}\sigma^2+ \mathrm{i}\chi_{3}\sigma^3).
\end{equation} Under time-reversal transformation,
$\dsZ_2^T: F \rightarrow F^\ast = (i\sigma^2) F (-i\sigma^2)$.

Two three-component vector operators can be conveniently
constructed with these edge Majorana operators ($a=1,2,3$):
\begin{equation}\label{eq: S=FF}
S^a =\frac{1}{2}\Tr F^\dagger\sigma^aF, \ \ \ \ K^a
=\frac{1}{2}\Tr F\sigma^aF^\dagger.
\end{equation} In fact, the boundary Majorana fermions have an
emergent SO(4) symmetry, and the two vectors correspond to the two
independent SU(2) subgroups of the SO(4). The full SO(4)
rotational symmetry among the four flavors of Majorana fermions is
decomposed to SU(2)$_\text{spin}\times$SU(2)$_\text{gauge}$,
generated by $\vect{S}$ and $\vect{K}$ respectively. For the
fermions in $F$, the SU(2)$_\text{spin}$ rotation corresponds to a
left rotation $F\to U^\dagger F$ with $U\in$SU(2)$_\text{spin}$,
while the SU(2)$_\text{gauge}$ rotation corresponds to a right
rotation $F\to FG$ with $G\in$SU(2)$_\text{gauge}$.

Under the constraint $\chi_{0}\chi_{1}\chi_{2}\chi_{3} = 1$, which
is equivalent to the requirement of gauge neutrality, \emph{i.e.}
$\vect{K}=0$. Therefore under the gauge constraint, the physical
state of the boundary is only two fold degenerate, and these
states are invariant under SU(2)$_\text{gauge}$. This means that
we are free to combine time-reversal symmetry with a
SU(2)$_\text{gauge}$ transformation. For example, we can define a
new time-reversal transformation $\mathcal{T}: F \rightarrow
F^\ast ( i\sigma^2) = - i\sigma^2 F $, this new time-reversal
transformation satisfies $\mathcal{T}^2 = -1$, and it is exactly
the same time-reversal transformation for spin-1/2 object. Thus we
conclude that under gauge constraint, four copies of Kitaev's
chain is equivalent to the Haldane's phase.

\section{Appendix B. Vison Loops in $^3$He B TSC}

In this appendix, we derive the effective theory along the vison
loop in the $^3$He B TSC. Let us start with \eqnref{he3}, and
first consider a straight vison line along the $x$-direction. The
vison line can be considered as a thin hollow cylinder through the
bulk of the TSC with a $\dsZ_2$ flux ($\pi$-flux) threading
through the hole of the tube. For this configuration, it could be
convenient to use the cylindrical coordinate defined as
$(x,y,z)=(x,\rho \cos \theta, \rho \sin \theta)$. Applying the
coordinate transform to \eqnref{he3}, the Sch\"ordinger equation
reads \beq \big(\ii\Gamma^1\partial_x+\ii\Gamma^2
e^{\Gamma^2\Gamma^3\theta}\partial_\rho+\ii\Gamma^3e^{\Gamma^2\Gamma^3\theta}\rho^{-1}(\partial_\theta-\ii
\omega_\theta)+m\Gamma^4\big)\chi = E \chi,\label{eq: vison eq}
\eeq where $\omega_\theta = \ii\Gamma^2\Gamma^3/2$ is the spin
connection that corresponds to threading the $\pi$-flux (as
$e^{\oint\ii\omega_\theta\dd\theta}=-1$). The low-energy fermion
modes around the vison line are given by the following ansatz in
the asymptotic limit, \beq \chi_a(x,\rho,\theta) \simeq e^{- m
\rho} e^{-\Gamma^2\Gamma^3\theta/2}\chi_a(x).\label{eq: vison sol}
\eeq Substitute \eqnref{eq: vison sol} to \eqnref{eq: vison eq},
one can see $\chi_a(x)$ must satisfy $\ii\Gamma^4\Gamma^2\chi_a(x)
= \chi_a(x)$ in order to obtain the low-energy modes (whose energy
$E\to0$ as the $x$-direction momentum $\ii\partial_x\to0$). The
matrix $\ii\Gamma^4\Gamma^2=\sigma^{31}$ has two eigenvectors of
the $+1$ eigenvalue: \beq
\chi_1=\frac{1}{\sqrt{2}}(1,1,0,0)^\intercal, \quad
\chi_2=\frac{1}{\sqrt{2}}(0,0,1,-1)^\intercal, \eeq corresponding
to the two counter-propagating Majorana modes along the vison
line. It is straight forward to see that the $4\times4$ matrix
$\Gamma^1=\sigma^{30}$ represented on the basis $(\chi_1,\chi_2)$
becomes the $2\times2$ matrix $\sigma^3$, so the effective $1d$
Hamiltonian should be $H_{1d,x}=\int\dd x
\chi^\intercal(x)\ii\sigma^3\partial_x\chi(x)$ as shown in
\eqnref{eq: vison Heff}.

In general, any operator $O$ (as a $4\times 4$ matrix) defined in
the $3d$ bulk can be thus projected to the subspace of the fermion
modes along the vison line, as the corresponding $2\times2$ matrix
$\tilde{O}$ by ($a,b=1,2$) \beq
\begin{split}
\tilde{O}_{ab} =& \int\dd\rho\,\dd\theta\,\chi_a^\intercal(x,\rho,\theta)O\chi_b(x,\rho,\theta) \\
 \simeq& \int\frac{\dd\theta}{2\pi}\, \chi_a^\intercal e^{\Gamma^2\Gamma^3\theta/2}O e^{-\Gamma^2\Gamma^3\theta/2}\chi_b.
\end{split}
\eeq In \tabref{tab: ops}, we conclude the projection of all
$4\times4$ Hermitian matrices (16 complete basis) to the
2-dimensional subspace of counter propagating Majorana modes along
the vison line. This establishes the correspondence between the
operators in the bulk and that on the vison line. One can see
$\Gamma^5$ in the bulk would correspond to $\sigma^2$ on the vison
line. So the action of the time-reversal symmetry $\dsZ_2^T$ is
reduced to $\chi\to\ii\sigma^2\chi$ on the vison line.

Given the effective Hamiltonian \eqnref{eq: vison Heff} and the
above $\dsZ_2^T$ symmetry on the vison line, it seems that if we
make even copies of the system, the vison line can be gapped out
by a bilinear mass term of the form $\chi^\intercal
\sigma^1\otimes A\chi$ (with $A=-A^\intercal$) which does not
breaks the time-reversal symmetry. However, this is only true for
our analysis of the straight vison line along the $x$-direction.
Because according to \tabref{tab: ops}, the mass term
$\chi^\intercal \sigma^1\otimes A\chi$ would extend to the bulk as
$\chi^\intercal \sigma^{13}\otimes A\chi$, which can not gap out
the vison lines along any other directions, as $\sigma^{13}$
commutes with both $\Gamma^2=\sigma^{10}$ and
$\Gamma^3=\sigma^{22}$. Therefore it is impossible to fully gap
out the vison loop by any fermion bilinear term.

\begin{table}[htdp]
\caption{Projection of bulk operators to the vison line ($x$-direction)}
\begin{center}
\begin{tabular}{rr}
$O$ & $\to\hspace{1cm}\tilde{O}$\\
\hline
$\sigma^{00}$ & $\int\frac{\dd\theta}{2\pi} \sigma^0=\sigma^0$ \\
$\sigma^{31}$ & $\int\frac{\dd\theta}{2\pi} \cos\theta\,\sigma^0=0$ \\
$\sigma^{03}$ & $-\int\frac{\dd\theta}{2\pi} \sin\theta\,\sigma^0=0$ \\
$\sigma^{13}$ & $\int\frac{\dd\theta}{2\pi} \sigma^1=\sigma^1$ \\
$\Gamma^3=\sigma^{22}$ & $\int\frac{\dd\theta}{2\pi} \cos\theta\,\sigma^1 = 0$\\
$\Gamma^2=\sigma^{10}$ & $-\int\frac{\dd\theta}{2\pi} \sin\theta\,\sigma^1 = 0$\\
$\Gamma^5=\sigma^{23}$ & $\int\frac{\dd\theta}{2\pi} \sigma^2=\sigma^2$\\
$\sigma^{12}$ & $-\int\frac{\dd\theta}{2\pi} \cos\theta\,\sigma^2=0$ \\
$\sigma^{20}$ & $-\int\frac{\dd\theta}{2\pi} \sin\theta\,\sigma^2=0$ \\
$\Gamma^1=\sigma^{30}$ & $\int\frac{\dd\theta}{2\pi} \sigma^3=\sigma^3$\\
$\sigma^{01}$ & $\int\frac{\dd\theta}{2\pi} \cos\theta\,\sigma^3=0$ \\
$\sigma^{33}$ & $-\int\frac{\dd\theta}{2\pi} \sin\theta\,\sigma^3=0$ \\
$\Gamma^4=\sigma^{21}$ & $0$\\
$\sigma^{02}$ & $0$\\
$\sigma^{11}$ & $0$\\
$\sigma^{32}$ & $0$\\
\end{tabular}
\end{center}
\label{tab: ops}
\end{table}


\end{document}